\documentclass[reprint,showpacs,showkeys,preprintnumbers,amsmath,amssymb,citeautoscript,aip,apl]{revtex4-2}
\usepackage{graphicx}
\usepackage{epstopdf}
\usepackage{color}
\usepackage{float}
\usepackage{soul}
\usepackage[hidelinks]{hyperref}

\begin{document}
\title{ Manipulation  of hybrid skyrmion dynamics by step DMI approach }
\author{Hitesh Chhabra}
\affiliation{School of Physical Sciences, National Institute of Science Education and Research, HBNI, Jatni-752050, India}
\author{Jayaseelan Dhakshinamoorthy}
\affiliation{School of Physical Sciences, National Institute of Science Education and Research, HBNI, Jatni-752050, India}
\author{Ajaya K. Nayak}
\email{ajaya@niser.ac.in}
\affiliation{School of Physical Sciences, National Institute of Science Education and Research, HBNI, Jatni-752050, India}

\date{\today}

\begin{abstract}
	
The dynamic behavior of non-collinear atomic spin arrangements in a topologically protected magnetic skyrmion plays a pivotal role in potential future spintronic technologies, including racetrack memory based ultra-high-density storage devices. However, the topological nature of the skyrmion comes with  an unwanted skyrmion Hall effect (SkHE) that poses a significant challenge in the practical application. Here we present a detailed micromagnetic simulation  study that delves into the controlled manipulation of skyrmion dynamics through a subtle engineering of Dzyaloshinskii-Moriya interaction (DMI) in a hybrid skyrmion racetrack. In particular, we introduce a gradient variation of bulk and interfacial DMI that results into a parabolic trajectory of Skyrmion Hall angle (SkHA), thereby allowing us to find a critical DMI ratio with almost zero SkHE. Most importantly,  we present a  novel approach involving engineering of a racetrack with strategically placed step DMI regions that gives us a meticulous control over the size and speed of the hybrid skyrmions. The present study gives a new direction for the simultaneous realization of stable skyrmions without SkHE and increased skyrmion speed  with optimized DMI engineering.  

\end{abstract}


\maketitle

   The  swirling kind of non-collinear  spin arrangement combined with a topological protected magnetic state in a skyrmion play important roles in the current driven dynamics required for several practical applications \cite{skylattice, realspace, electrodynamics, repulsion, current-drivenRT, isolated}. The ultra-small nanometric size of the skyrmions and their lower depinning current density compared to the contemporary systems make a genuine case for the low energy consumption based  ultra-high density storage devices \cite{small, racetrackmemory, lowcurrent, RTlowI}. Furthermore, their stable topological structure offers immunity against minor distortions  benefiting in preserving the stored information and hence are envisioned to be employed as non-volatile memory and logic bits in spintronic technologies \cite{constrictedgeometries, impurities, isolated, benefits}. However, the topological nature of the skyrmions albeit an advantage also leads to a detrimental skyrmion Hall effect (SkHE), analogous to the magnus force resulting from the rotational and the translational motion of the skyrmion due to torque exerted on it by the driving current  \cite{torque, constrictedgeometries, SkyHE}. This unwanted SkHE leads to the transverse motion of the skyrmions, resulting in their  annihilation at the racetrack edges. 
   
   Various approaches have been made to combat the SkHE issue by providing an equal and opposite transvese motion of skyrmion in a bilayer stack of perpendicularly magnetized ferromagnets coupled via antiferromagnetic exchange in synthetic antiferromagnets (SyAF) \cite{AFMSky, bilayer, HMS, Synthetic}. Here the two layers of ferromagnets show skyrmions of opposite topological charge producing a coupled structure experiencing a counter-balanced SkHE. Similar concept is also proposed for the skyrmions in multilayer ferrimagnets of GdFeCo films with compensated angular momentum at a critical temperature \cite{ferriSky, ferriSky2}. Alternate approach includes creating anisotropy gradients or guiding the skyrmions along predefined path by creating potential well between the boundaries \cite{edge, 1D}. Likewise, by light ion irradiation, the energy landscape of a skyrmion based racetrack can be engineered to force the skyrmion to follow a predetermined route by using repulsive forces from the side walls \cite{ion}.
   
   Another important approach involves helicity manipulation in hybrid skyrmions \cite{hyb, HMS}. The non-trivial chiral pin structure of the skyrmions originates from the competing  Heisenberg exchange and Dzyaloshinskii-Moriya interaction (DMI) in a broken inversion symmetry system \cite{Dzyaloshinskii, Moriya, Bogdanovchirality}. Depending on the nature of the DMI, different kinds of spin rotations (chirality and helicity) can be stabilized. For instance, Bloch skyrmions observed in most of the bulk DMI materials exhibit spin helicity of $\pm$$ \pi/2 $, whereas Neel skyrmions, mostly found in interfacial DMI systems, display helicity of $ 0 / \pi $ \cite{topological, applicationreview}. By utilizing the Neel and Bloch type spin helicities, it is possible to realize a composite skyrmion structure, which can be driven without transverse motion or SkHE. These hybrid skyrmionic topological structures are promising candidates as they can be stabilized in ferromagnetic materials itself. A recent report also discusses the possibility of stabilizing hybrid skyrmion by the interplay of dipolar interaction and interfacial DMI (iDMI) \cite{hybstability}. In addition to restricting the SkHE, helicity control also offers other advantages, e.g. self-selectivity in a multiple racetrack geometry for logic operations \cite{logicgate}. The helicity  offers a significant advantage in spintronics applications, as it can be easily manipulated and controlled by DMI engineering, voltage/electric field control \cite{Vcontrol, EFcontrol} or via strain  \cite{Scontrol}. 
   
  The present work  focuses on the DMI engineered hybrid skyrmions, which can be leveraged to achieve faster skyrmion dynamics for practical device implementation. In particular, we look into the trajectory of hybrid skyrmions under a gradient/step  of bulk and  interfacial DMI, which can be instrumental for the understanding of skyrmion dynamics in real materials.  The obtained trajectory uncovers the direct dependence of skyrmion size, speed, SkHE, and topology on the underlying mixed DMIs.  By doing so, we here suggest a way to improve the skyrmion dynamics via DMI ratio tuning for the case of hybrid skyrmion moving without skyrmion Hall angle. Our work utilizes skyrmion distortion in a controllable way based on the trajectory to increase its speed.

{\bf Simulation details:} All the micromagnetic simulations are performed using the Object Oriented MicroMagnetic Framework  (OOMMF) code \cite{soft}.  Our basic geometry includes coupling two ferromagnetic layers via ferromagnetic exchange interaction. The simulation dimension of the slab geometry is taken as 1250 nm $\times$ 500 nm $\times$ 1.2 nm with the mesh size of 2 nm  $\times$ 2 nm $\times$ 0.6 nm.  For the purpose of hybrid DMI in the system, bulk DMI is given to the top layer while the bottom layer consists of interfacial DMI originating due to the underlying heavy metal, also used to provide spin orbit current. The material parameters used for the simulations consist of saturation magnetization ($ M_S $) 5.8 $\times$ 10$^{5}$ A/m, uniaxial magnetic anisotropy of 6 $\times$ 10$^{5}$ MJ/m$^{3}$ in the z-direction, intralayer exchange constant of 15 pJ, interlayer exchange constant of 35 pJ, and the gilbert damping constant as 1.  The magnitude of DMI is varied as per the requirement. We use two kinds of structure in our approach as described below.


      In the first case, the trajectory of hybrid skyrmion is studied under a gradient of bulk and Néel DMI in a bilayer structure. The magnitude of bulk DMI on the top layer is gradually decreased  while the interfacial DMI strength on the bottom layer is gradually increased by moving from one end  to other end of the stack. The skyrmion is driven by the help of spin current and its dynamics is studied with respect to DMI ratio (i.e., DMI$_{\text{Bulk}}$/DMI$_{\text{Néel}}$).

      In the second case, we design a zero SkHE racetrack for hybrid skyrmions where both the layers with uniform bulk and Néel DMI are set to a critical DMI ratio obtained from the first case. In order to improve the velocity of the hybrid skyrmion, we introduce a few regions with high bulk DMI magnitude, on the top layer of the track. These regions can be created by doing selective light or heavy ion irradiation method or by other means of DMI control such as voltage or strain \cite{Vcontrol, EFcontrol, Scontrol}. We can also produce similar and even improved zero SKHE racetracks instead by changing the Néel DMI in the bottom layer.
\begin{figure}[h!]
	\includegraphics[width = 9cm, height = 6.5cm, clip]{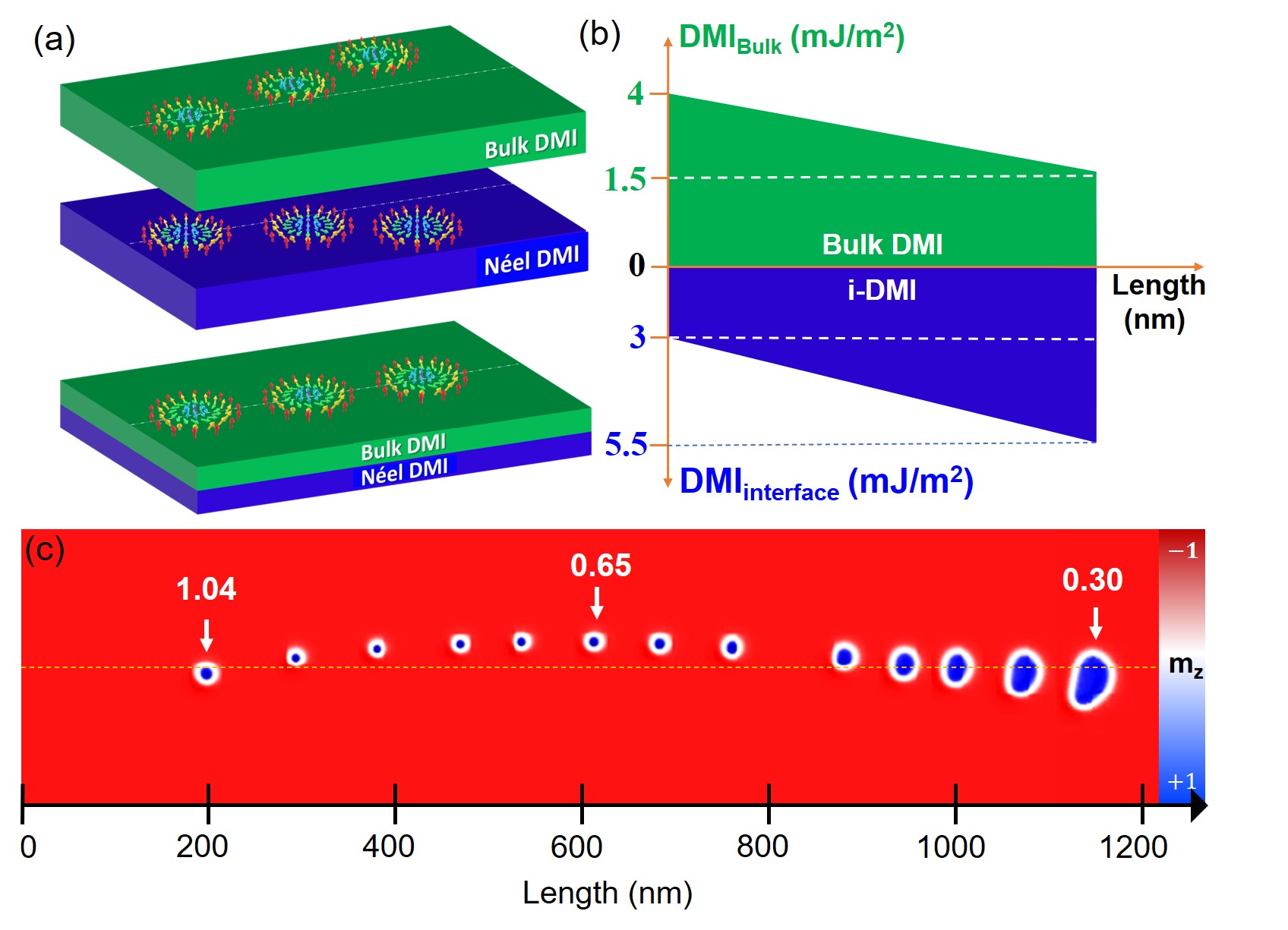}
	\caption{\label{fig1} (a) Schematics of combining contrasting SkHE using bulk and Néel-type skyrmions to obtain zero SkHE. (b) Along the proposed track, bulk DMI decreases in the top layer while the Néel DMI increases in the bottom layer. (c) Resultant parabolic trajectory of the hybrid skyrmion in a track with opposite gradient of DMI. Color bar shows the magnetization state.}
\label{fig1}	
\end{figure}

{\bf Tuning the Skyrmion Hall Effect to zero:} We begin with constructing a layered thin film geometry to study how a gradient DMI  affects the motion of a hybrid skyrmion for which the bulk DMI decreases from 4 mJ/m$^{2}$ to 1.5  mJ/m$^{2}$ while the Néel increases from 3 mJ/m$^{2}$ to 5.5 mJ/m$^{2}$ by changing the ratio of overall DMI$_{\text{Bulk}}$:DMI$_{\text{Néel}}$ from 1.33 at one end to 0.27 at the other end, as depicted in the Fig.\ref{fig1}(b). Under this scenario, we find a continuous change in the skyrmion Hall angle (SkHA) producing a parabolic trajectory as shown in Fig.\ref{fig1}(c). This is in contrast to the observation of angled rectilinear motion  normally seen in the case of a track with uniform bulk or interfacial DMI. At the starting point of the track, the bulk DMI dominates over the Néel DMI, resulting the skyrmion gets deflected  towards the left edge of the sample. On the contrary, SkHE is observed to be directed towards the right edge of the sample in areas where the Néel DMI dominates over the bulk one. Thus,  tuning the DMI ratio decides the direction of SkHE which can be nullified by controlling the helicity of the hybrid skyrmion as shown in Fig\ref{fig1}(a). 

Next, we concentrate on the velocity of hybrid skyrmion at different hybrid DMI ratio, as depicted in Fig.\ref{fig2}(a). The x-direction velocity is found to be about 650 m/s  for a DMI ratio of 0.3.  The velocity starts decreasing before reaching a a minima for the DMI ratio of 0.7. Interestingly, at this DMI ratio the y-direction velocity (hence the SkHE) is also found to be zero, resulting the net velocity is because of the motion of the skyrmion in the x-direction [see Fig.\ref{fig2}(b)].  This can be summarized as DMI ratio greater than 0.7 leads to the SkHE in the +y direction, while ratio less than 0.6 leads to SkHE in -y direction, as shown in Fig\ref{fig1}(c).  It can also be seen that the  appearance of SkHE is accompanied by a continuous variation in the size of skyrmion, which becomes larger towards the ends of racterack. This indicates that any larger value either of bulk or Néel DMI increases the size of  skyrmion. On the other hand, a comparable value of  bulk and Néel DMI  forms a highly stable and small size (diameter $\approx$ 28 nm) hybrid skyrmion. In these region the DMI ratio is found to be in the range of 0.6 to 0.7, which can be considered as the critical ratio to obtain almost zero SkHE with stable skyrmion dynamics.

\begin{figure}[h!]
	\includegraphics[width = 9.5cm, height = 8.5cm, clip]{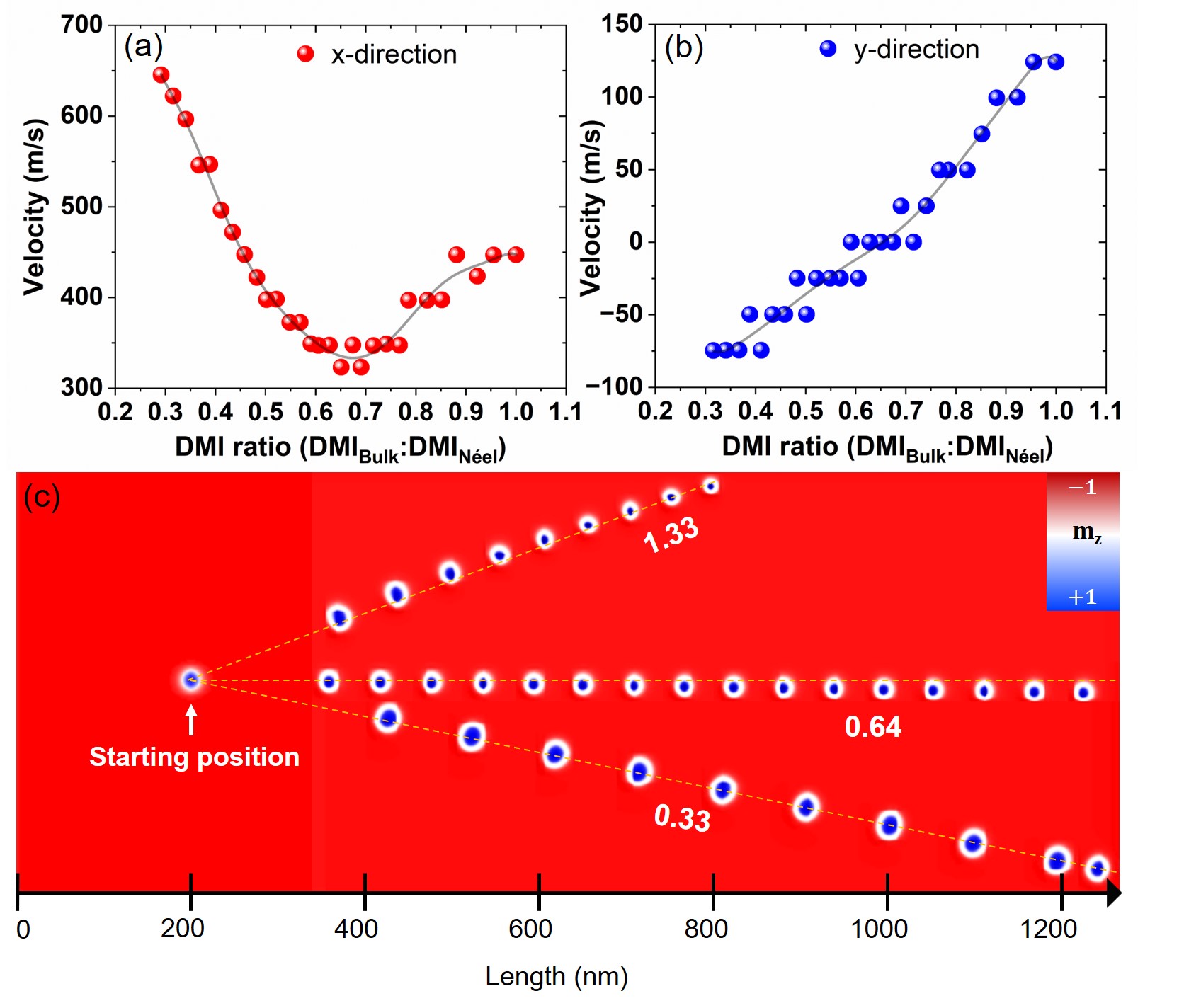}
	\caption{\label{fig2}  DMI ratio dependent skyrmion velocity in the (a) $x$ direction and (b) $y$ direction. The lines are guide to the eye. (c) Variation of SkHE with three different DMI ratio. The color bar shows the magnetization state. }
\label{fig2}	
\end{figure}


As discussed above, a large skyrmion velocity can be obtained when the DMI ratio is too less or too high  as compared to the region sustaining zero SkHE. The larger size of hybrid skyrmion in these regions is primarily responsible for the increase in the velocity \cite{sizedependence}. However, these deformed topological structures at the extreme DMI ratios are unstable and quickly get  annihilated. Although they offer very high velocity, the presence of significant SkHE and their proneness to distortion is an obstacle in real applications. The distortion as well as the increase in the velocity is found to be more effective in the D$_{\text{Néel}}$ dominating regions compared to that of D$_{\text{Bulk}}$. Based on these observations, it is possible to increase the velocity of hybrid skyrmion by engineering the DMI magnitude in  both of the  layers. Hence in the following section, the bulk DMI in certain region of the top layer  is increased while  Néel DMI of the bottom layer is kept uniform throughout the track keeping in mind the experimental feasibility. It is important to note here that similar and even an improved dynamics can also be achieved by changing Néel DMI instead. 


{\bf Tailoring the Skyrmion Dynamics:} Despite of achieving a streamline flow of skyrmion without SkHE, the velocity of skyrmion is found to be compromised at the critical DMI ratio. In order to improve the velocity of hybrid skyrmion, we engineer the hybrid racetrack (i.e.,  DMI$_{\text{Bulk}}$:DMI$_{\text{Néel}}$) by introducing steps with higher DMI$_{\text{Bulk}}$ region on the top layer, as  shown in Fig. \ref{fig3}(a). The difference in DMI magnitude between the engineered region and rest of the racetrack is denoted as $\Delta$DMI$_{bulk}$. The width of the strip ranges from 10 to 40 nm, whereas the $\Delta$DMI$_{bulk}$ varies from 0.1 to 0.8 mJ/m$^{2}$. Additionally, the DMI ratio at the strip region changes from 0.66 to 0.82. The hybrid skyrmion is allowed to travel 250 nm with a velocity of ~350 m/s before encountering any DMI change to ensure its stability. The velocity of the hybrid skyrmion after encountering strips of different width and  magnitude of $\Delta$DMI$_{bulk}$ is represented in  Fig. \ref{fig3}(b). As anticipated, an increase in the velocity can be realized as the magnitude of $\Delta$DMI$_{bulk}$ increases within the constant-width strip region. Likewise, a comparable increment in the velocity is also noticed when the strip width of modified region is increased while $\Delta$DMI$_{bulk}$ is kept constant. By altering either one or both of these factors, it becomes evident that the hybrid skyrmion can attain a higher velocity, potentially reaching speeds of up to 450 m/s. We also expect that the speed can be further increased with optimal DMI ratio and other parameters in a real material.


\begin{figure}[tb!]
	\includegraphics[width = 9cm, height = 10.5cm, clip]{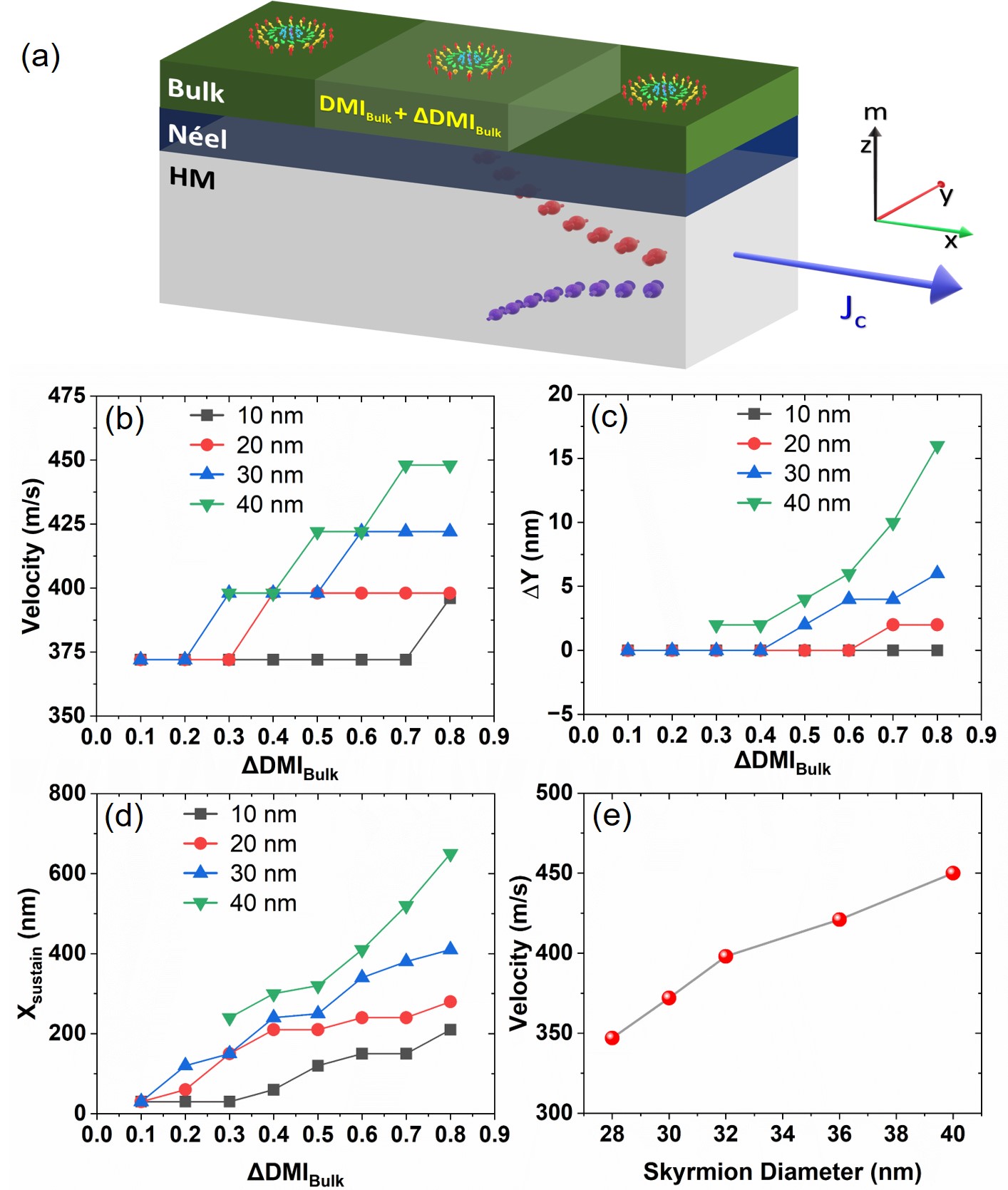}
	\caption{\label{fig3} (a) Schematic representation of introducing a step DMI in the racetrack. The highlighted area (in the topmost layer) defines a strip of length ranging from 10 nm to 40 nm having an increased bulk DMI from 0.1-0.8 mJ/m$^{2}$.  Depending on the width and magnitude of $\Delta$DMI$_{bulk}$ (b) velocity of hybrid skyrmion in $x$ direction, (c) deviation from the skyrmion in $y$ direction and (d) distance in $x$  over which an increase in velocity of the skyrmion is sustained.  (e) Dependence of skyrmion velocity on the skyrmion size.}
\label{fig3}	
\end{figure}


Although the modified DMI regions help in increasing the skyrmion velocity, it can also deliberately increase the size of the hybrid skyrmion, consequently leading to the appearance of SkHE. Figure \ref{fig3}(c) shows the deviation of hybrid skyrmion in the y-direction with different  $\Delta$DMI$_{bulk}$ for variable strip widths. Excessive DMI increase in a wider strip leads to substantial SkHE, which should be minimized to maintain the stability of skyrmion's trajectory. Furthermore it is also found that the change in  skyrmion topology persists over a distance exceeding the dimension of the engineered strip, resulting in an increased velocity as shown in Fig. \ref{fig3}(d). This sustenance of increased velocity is an interesting finding that we have leveraged in the following section.

While  maintaining an elevated velocity, it is observed that the size of the skyrmion exceeds than that found in the typical non-engineered track. However, after traveling a certain distance the skyrmion reverts  to its original size. As it is previously mentioned, the increase in the size of hybrid skyrmion directly correlates with the rise in its velocity. This direct relationship between skyrmion velocity and size is depicted in Fig. \ref{fig3}(e). Specifically, an increase of 12 nanometers in the skyrmion diameter results in a velocity rise of 100 m/s. 


\begin{figure}[tb!]
	\includegraphics[width = 9cm, height = 7.5cm, clip]{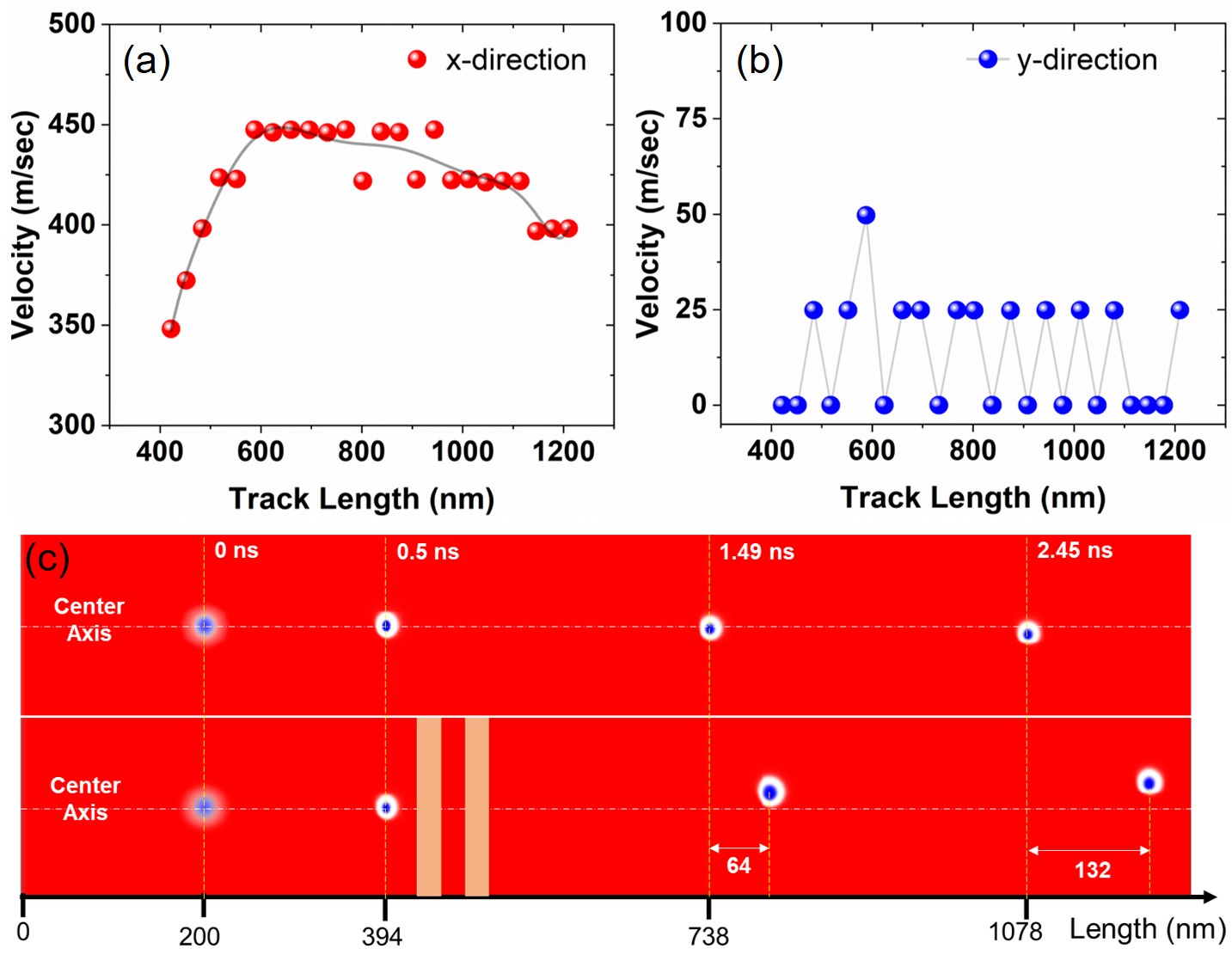}
	\caption{\label{fig4} Recorded velocity in the (a) x and (b) y directions plotted against the track length for a $\mu$m length racetrack with two optimized strips. (c) Comparison of skyrmion dynamics in regular (upper) and modified (lower) racetrack. The highlighted vertical strips represent the engineered region.}
	\label{fig4}	
\end{figure}

Similarly, we find the optimal $\Delta$DMI$_{bulk}$ magnitude, strip width and their spacing that would allow the hybrid skyrmion structure to travel about a micrometer length of distance with highest possible velocity while experiencing minimal SkHE. Based on the dynamics study illustrated in Fig. \ref{fig3}(b-e), the ideal strip width is determined to be 40 nm, in conjunction with a $\Delta$DMI$_{bulk}$ value of 0.5mJ/m$^{2}$ and the optimal spacing of 40 nm. As shown in  Fig. \ref{fig4}(a), it is evident that the velocity of skyrmion starts to increase upon reaching the edge of strip and maintains an increase of 100 m/s for approximately 500 nm. The velocity in y-direction due to the slight skyrmion distortion is plotted in Fig. \ref{fig4}(b), which shows a velocity magnitude ($\sim$25 m/s) much lesser than that of the velocity along the x-direction ($\sim$410m/s). In a conventional racetrack, which is approximately 1$\mu$m in length, a hybrid skyrmion without any modification, takes 3.59 ns to traverse the entire length. The deviation from the current direction is found to be 12 nm  which gives a SkHA of $\sim0.7^{\circ}$.  However, in a racetrack with engineered strip DMI, the operation is completed in 2.93 ns, and the deviation of skyrmion from the x-direction increases slightly to 26 nm that results in a SkHA of $\sim1.4^{\circ}$. The comparison of skyrmion dynamics with conventional and engineered racetracks are depicted in Fig. \ref{fig4}(c), where the bottom track with highlighted region represents the strips with DMI change. Thus, by having optimum strip separation of increased magnitude of DMI$_{\text{Bulk}}$ or DMI$_{\text{Néel}}$, it is possible to enhance the velocity of hybrid skyrmion while ensuring minimum displacement from the zero SkHE trajectory.


 The step DMI approach in the present study offers dual advantages in hybrid skyrmion systems. Firstly, it enhances the skyrmion speed and therefore reduces the computation time. Secondly, it enables DMI dependent guidance of skyrmions movement to a particular direction by providing an overall SkHA.  Through multiple strips of $\Delta$DMI$_{bulk}$ and optimum spacing between them, precise control of current-driven hybrid skyrmions can be achieved which can then be aligned to designated lanes as required. Hence the present study makes a direct impact on the designing of skyrmion-based logic devices \cite{logicgate}. Experimentally, the step DMI can be easily designed by  exposing the track to ion beam irradiation through proper masking. The ion beam can easily change the local magnetic anisotropy at the exposed area and thereby creating the required DMI step.


	To conclude, hybrid bilayer structure  provides a unique way to fine tune the DMI, which offers the capability to govern the helicity of hybrid skyrmion such that it does not induce any transverse motion. The obtained parabolic trajectory of the skyrmion in case of hybrid DMI reveals a way to increase the velocity of hybrid skyrmion travelling in a straight line through selective DMI engineering by introducing strips of higher DMI$_{\text{Bulk}}$ or DMI$_{\text{Néel}}$. From the comprehensive studies on the impact of altering the bulk DMI, we have meticulously fine-tuned the strip parameters to achieve the highest possible skyrmion velocity over an extended distances without skyrmion deformation. Overall, a skyrmion based racetrack of approximately 1$\mu$m in length can be implemented in the device application where step DMI can be introduced  using techniques such as ion irradiation or electrical/strain control of DMI. This modification can enhance the speed of skyrmion by about 30\% and hence reduces the operation time.


\section*{}
AKN acknowledges the support from Department of Atomic Energy (DAE) and SERB research grant (CRG/2022/002517) of the Government of India for financial support.

\section*{AUTHOR DECLARATIONS}
\textbf{Conflict of Interest}\\
The authors have no conflicts to disclose.

\section*{Data Availability Statement}

The data that support the findings of this study are available from the corresponding author upon reasonable request.


\end{document}